\documentclass[twocolumn,showpacs,preprintnumbers,prl,amsmath,amssymb]{revtex4}

\usepackage{graphicx}
\usepackage{dcolumn}
\usepackage{bm}

\begin{document}

\preprint{APS/123-QED}

\title{Polygonization of carbon nanotubes}

\author{Dmitry Golovaty}
\email{dmitry@math.uakron.edu}
\author{Shannon Talbott}%
\affiliation{%
Department of Theoretical and Applied Mathematics, The University of Akron, Akron, OH 44325-4002
}%

\date{\today}

\begin{abstract}
We use a multiscale procedure to derive a simple continuum model of multiwalled carbon nanotubes that takes into account both strong covalent bonds within graphene layers and weak bonds between atoms in different layers. The model predicts polygonization of crossections of large multiwalled nanotubes as a consequence of their curvature-induced turbostratic structure.
\end{abstract}

\maketitle


Polygonization of crosssections of large multiwalled carbon nanotubes (MWNTs), especially following heat treatment, has been observed in a number of studies \cite{kiang,wu_cheng,yoon}.
The physics of this phenomenon is generally well-understood---the curvature-induced mismatch between the lattices of the adjacent graphene shells can be mollified by flattening the shells at the expense of creating the line defects needed to maintain the cylindrical shape of the tube \cite{yoon}.

The modeling of polygonization have proved more elusive: the atomistic modeling of large tubes is computationally expensive, while continuum models that do not take into account atomic structure of graphene are inadequate to reproduce lattice-dependent features of MWNTs \cite{yakobson,arroyo}. To our knowledge, the only existing model \cite{yoon} of polygonization in MWNTs is rather complex and it assumes from the start that a nanotube is polygonal. The extensive first-principle calculations are then used to compute the energy of the defects and the planar sections of the tube in order to select the optimal shape of MWNT. 

In this letter we employ a continuum theory derived by upscaling a simple atomistic model to predict polygonization of large MWNTs. Our main premise is that polygonization is the result of a competition between the two types of interatomic interactions---the interactions within individual graphene layers and the interactions between these layers---and it is controlled by the diameter but not the number of walls in the MWNT. We take advantage of the last observation by considering the simplest case of a two-walled carbon nanotube.

To derive a continuum model of a two-walled nanotube we will assume that the diameter of each tube is much larger than the length of an interatomic bond in a hexagonal carbon lattice. Then a {\em{microscale}} object has dimensions of order of the carbon bond length and a {\em{macroscale}} object has dimensions of order of the diameter of the nanotube.

\begin{figure}[ht]
\includegraphics[width=2in]{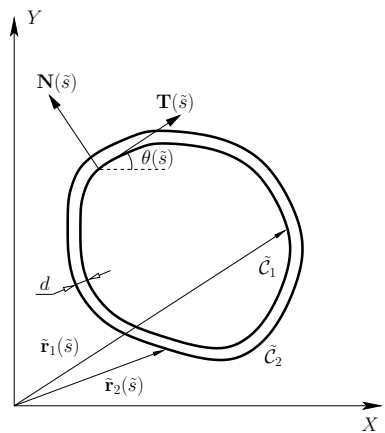}
\caption{\label{fig:crossection} Crossection of a two-walled carbon nanotube.}
\end{figure}

\smallskip
\noindent {\bf{Macroscopic geometry:}} Suppose that two smooth, closed concentric curves $\tilde{\mathcal C}_1$ and $\tilde{\mathcal C}_2$ represent a crossection of a two-walled carbon nanotube by a plane perpendicular to the axis of the tube (Fig. \ref{fig:crossection}). To simplify mathematics we will assume that these curves are parallel, that is the distance between $\tilde{\mathcal C}_1$ and $\tilde{\mathcal C}_2$ when measured along any normal to $\tilde{\mathcal C}_1$ is equal to the same constant value of $d$. The $\mathbf R^2$-valued function $\tilde{\mathbf r}_1(\tilde{s})$ parametrizes $\tilde{\mathcal C}_1$ with respect to its arclength $\tilde{s}\in[0,L]$ measured counterclockwise from some fixed initial point $\tilde{\mathbf r(0)}$. Here $L$ is the length of $\tilde{\mathcal C}_1$. 

An orthogonal frame at $\tilde{\mathbf r}_1(\tilde{s})$ is given by a pair $(\mathbf T(\tilde{s}),\mathbf N(\tilde{s}))$ with $\mathbf T(\tilde{s})=-\tilde{\mathbf r}_1^\prime(\tilde{s})=\left<\cos{\theta(\tilde{s})},\sin{\theta(\tilde{s})}\right>$ and $\mathbf N(\tilde{s})=\left<-\sin{\theta(\tilde{s})},\cos{\theta(\tilde{s})}\right>$, where $\theta(\tilde{s})$ is the angle between $\mathbf T(\tilde{s})$ and the $x-$axis
at the point $\tilde{\mathbf r}_1(\tilde{s})$. The curvature of $\tilde{\mathcal C}_1$ can be computed via the relation $\kappa(\tilde{s})=\theta^\prime(\tilde{s})$ then, according to the Frenet formulas, $\mathbf T^\prime(\tilde{s})=\kappa(\tilde{s})\mathbf N(\tilde{s})$ and $\mathbf N^\prime(\tilde{s})=-\kappa(\tilde{s})\mathbf T(\tilde{s})$.

In what follows, we will require that $d\kappa(\tilde{s})<1$ for all $\tilde{s}\in[0,L]$. This assumption is physically reasonable as it automatically holds for any nanotube with a convex crossection and it only rules out relatively large inward folds of the curve $\tilde{\mathcal C}_1$. Then we can parametrize $\tilde{\mathcal C}_2$ with respect to $\tilde{s}$ by setting $\tilde{\mathbf r}_2(\tilde{s})=\tilde{\mathbf r}_1(\tilde{s})+d\mathbf N(\tilde{s})$. Note that $\tilde{s}$ is not an arclength parameter for $\tilde{\mathbf r}_2$. Indeed, $\tilde{\mathbf r}^\prime_2(\tilde{s})=-\left(1+d\kappa(\tilde{s})\right)\mathbf T(\tilde{s})$ and the distance traveled along $\tilde{\mathcal C}_2$ from $\tilde{\mathbf r}_2(0)$ to $\tilde{\mathbf r}_2(s)$ is $l(\tilde{s})=\int_0^{\tilde{s}}(1+d\kappa(\sigma))d\sigma=\tilde{s}+(\theta(\tilde{s})-\theta(0))d$ while the distance traveled along $\tilde{\mathcal C}_1$ from $\tilde{\mathbf r}_1(0)$ to $\tilde{\mathbf r}_1(s)$ is equal to $\tilde{s}$. Then the running difference between the distances traveled along  $\tilde{\mathcal C}_2$ and $\tilde{\mathcal C}_1$ is $\mathcal L(\tilde{s})=l(\tilde{s})-\tilde{s}=d(\theta(\tilde{s})-\theta(0))$ and the overall difference between the lengths of the two curves is $\mathcal L(L)=2\pi d$. Further, by an appropriate rotation of coordinates, we can set $\theta(0)=0$ to obtain
\begin{equation}
\label{eq:1}
\mathcal L(\tilde{s})=d\theta(\tilde{s}).
\end{equation}

\smallskip
\noindent {\bf{Microscopic structure:}} Having established the macroscopic geometrical framework for our model we need to endow it with the atomic structure of carbon. Since for mathematical simplicity we operate with crossections of nanotubes that are intrinsically one-dimensional while the hexagonal atomic lattice of graphene is two-dimensional, we will assume that the projection of the lattice on each crossection is a one-dimensional chain of atoms that retains the important physical characteristics of the original lattice. Although this assumption might not lead to a quantitatively accurate effective continuum model, we expect that the effective model based on physically reasonable assumptions should be qualitatively accurate.

\begin{figure}[ht]
\includegraphics[width=2.5in]{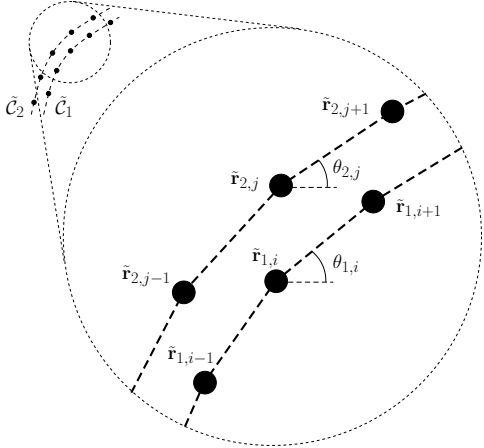}
\caption{\label{fig:atomic} Positions of atoms on the curves $\mathcal C_1$ and $\mathcal C_2$.}
\end{figure}

Suppose that two sets (chains) of equidistant carbon atoms are imbedded in the curves $\tilde{\mathcal C}_1$ and $\tilde{\mathcal C}_2$, respectively, as shown in Fig. \ref{fig:atomic}. The distance between the neighboring atoms in each chain will be fixed to the same constant $h$ to reflect the fact that the atoms within a graphene layer are connected by essentially inextensible strong $sp^2\ \sigma$-bonds. Then the curve $\tilde{\mathcal C}_2$ contains $n=\frac{2\pi d}{h}$ more atoms that the curve $\tilde{\mathcal C}_1$.

\smallskip
\noindent {\bf{Energy:}} We will assume that the overall energy of the two-curve-system consists of two parts: the energy due to bending of the adjacent bonds and the energy due to weak delocalized $\pi$-bonds between the atoms imbedded in $\tilde{\mathcal C}_1$ and $\tilde{\mathcal C}_2$, respectively. 

Suppose that the energy associated with two bonds joined at an angle $\theta$ is $f(\theta)$ where $f:(0,2\pi)\to \mathbf R$ is a smooth convex nondimensional function that has a minimum value of $0$ at $\theta=\pi$ and satisfies $f(\pi-\theta)=f(\pi+\theta)$. Further, let the weak interaction between the atoms in $\tilde{\mathcal C}_1$ and $\tilde{\mathcal C}_2$ be described by a Lennard-Jones-type potential $g(r)={\left(\frac{1}{r}\right)}^{12}-{\left(\frac{1}{r}\right)}^6$. Then the total energy of the system is
\begin{eqnarray*}
&\tilde E=\tilde\alpha\sum_{i=1}^{N-1}f\left(\pi+\theta_{1,i}-\theta_{1,i+1}\right) & \\
&+\tilde\alpha\sum_{j=1}^{N+n-1}f\left(\pi+\theta_{1,j}-\theta_{1,j+1}\right) & \\
&+\tilde\gamma\sum_{i=1}^{N-1}\sum_{j=1}^{N+n-1}g\left(\frac{\left|\tilde{\mathbf r}_{1,i}-\tilde{\mathbf r}_{2,j}\right|}{d}\right),&
\end{eqnarray*}
where $\tilde\alpha$ and $\tilde\gamma$ are the dimensional scaling factors and $N$ is the number of atoms imbedded in $\tilde{\mathcal C}_1$. 

From now on we will assume that $d,h\ll L$ and $n=\frac{2\pi d}{h}>2$ is a fixed positive integer. Nondimesionalizing the variables $\mathbf r=\tilde{\mathbf r}/L$, $s=\tilde{s}/L$ and rescaling $\tilde E$ by $\tilde \gamma$, the nondimensional energy is
\begin{eqnarray}
E&=&\alpha\sum_{i=1}^{N-1}f\left(\pi+\theta_{1,i}-\theta_{1,i+1}\right) \nonumber \\ 
&+&\alpha\sum_{j=1}^{N+n-1}f\left(\pi+\theta_{1,j}-\theta_{1,j+1}\right) \\
&+&\sum_{i=1}^{N-1}\sum_{j=1}^{N+n-1}g\left(\frac{\left|{\mathbf r}_{1,i}-{\mathbf r}_{2,j}\right|}{\delta}\right), \nonumber
\label{eq:2}
\end{eqnarray}
where $\delta=d/L\ll 1$, $\epsilon=h/L\ll 1$, and $\alpha=\tilde\alpha/\tilde \gamma$ are the nondimensional parameters of the system. Further, we will set $\mathcal C_1$ and $\mathcal C_2$ to represent $\tilde{\mathcal C}_1$ and $\tilde{\mathcal C}_2$ in nondimensional variables.

\smallskip
\noindent{\bf{Effective model:}} Using the smoothness of $\mathbf r_1$ we have that $\theta_{1,i+1}-\theta_{1,i}=\theta^\prime(s_i)\epsilon+o\left(\epsilon\right)$ for every $i=1,\ldots,N$, where $s_i$ denotes the position of the $i$-th atom on the curve $\mathcal C_1$. Since $f$ is smooth, the symmetry of $f$ and the fact that it has a minimum at $\theta=\pi$ imply that $f\left(\pi+\theta_{1,i}-\theta_{1,i+1}\right)=\frac{1}{2}f^{\prime\prime}(\pi){\left(\theta^\prime(s_i)\right)}^2\epsilon^2+o(\epsilon^2)$ for every $i=1,\ldots,N$. Then
\begin{eqnarray}
& \alpha\sum_{i=1}^{N-1}f\left(\pi+\theta_{1,i}-\theta_{1,i+1}\right) & \nonumber \\  
& =\alpha \epsilon\left[\sum_{i=1}^{N-1}\frac{1}{2}f^{\prime\prime}(\pi){\left(\theta^\prime(s_i)\right)}^2\epsilon\right]+o(\epsilon) & \\ 
&=\frac{\alpha\epsilon f^{\prime\prime}(\pi)}{2}\int_0^1{\left(\theta^\prime\right)}^2ds+o(\epsilon). & \nonumber
\label{eq:3}
\end{eqnarray}
Since exactly the same argument applies to the analogous sum over the curve $\mathcal C_2$, we have that the leading contribution to the total energy due to bending of interatomic bonds is 
\begin{equation}
\label{eq:4}
E_b=\mu\epsilon\int_0^1{\left(\theta^\prime\right)}^2ds,
\end{equation}
where $\mu=\alpha f^{\prime\prime}(\pi)$. This expression (cf. e.g. \cite{arroyo}
) corresponds to the well-known Euler elastica model for long slender beams \cite{antman}.

Next we develop the effective expression for the weak interaction energy. Fix an atom $i$ on an inner curve $\mathcal C_1$ and let $A=O\left(\epsilon^{1/2}\right)$ be an arbitrary number that we can use as a "mesoscale" because $\epsilon\ll A\ll 1$ . Set $\Lambda_A$ to be the set of all atoms on the curve $\mathcal C_2$ that are closer than $A$ to the atom $i$. Then \(\sum_{j=1}^{N+n-1}g\left(\frac{\left|\mathbf r_{1,i}-\mathbf r_{2,j}\right|}{\delta}\right)=\sum_{j\in\Lambda_A}g\left(\frac{\left|\mathbf r_{1,i}-\mathbf r_{2,j}\right|}{\delta}\right)+\sum_{j\notin\Lambda_A}g\left(\frac{\left|\mathbf r_{1,i}-\mathbf r_{2,j}\right|}{\delta}\right)\). The second sum can be approximated as follows
\begin{equation}
\label{eq:4.5}
\sum_{j\notin\Lambda_A}g\left(\frac{\left|\mathbf r_{1,i}-\mathbf r_{2,j}\right|}{\delta}\right)\sim {\left(\frac{\delta}{A}\right)}^6\frac{1}{\epsilon}=o\left(\epsilon^2\right),
\end{equation}
because 
\begin{equation}
\label{eq:4.6}
{\delta}/{\epsilon}={d}/{h}={n}/{2\pi}.
\end{equation}

\begin{figure}[ht]
\includegraphics[width=3in]{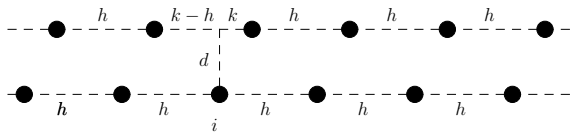}
\caption{\label{fig:chain} Two offset chains of atoms.}
\end{figure}

Having estimated the contribution to the weak interaction energy from the atoms on $\mathcal C_2$ that are macroscopically distant from the atom $i\in\mathcal C_1$, we now estimate the contribution to the energy due to the atoms that are microscopically close to the atom $i$. Since the curve $\mathcal C_2$ is smooth and since $A\ll 1$, the part of the curve $\mathcal C_2$ that is closer than $A$ to the atom $i$ can be approximated by the tangent line to $\mathcal C_2$ at the point $s_i$. That is, locally, we have a situation depicted in Fig. \ref{fig:chain}---in the macroscopically small neighborhood of the atom $i$ the curves appear as two infinite straight lines of atoms offset by some distance $k$ when viewed from the microscopic perspective.
Therefore the energy of weak interaction between the atom $i$ and the atoms in $\Lambda_A$ is 
\begin{eqnarray*}
\label{eq:5}
\sum_{j\in\Lambda_A}g\left(\frac{\left|\mathbf r_{1,i}-\mathbf r_{2,j}\right|}{\delta}\right) \sim\sum_{j=-\infty}^\infty g\left(\frac{\sqrt{{(\epsilon j+k)}^2+\delta^2}}{\delta}\right),
\end{eqnarray*}
to the leading order in $\epsilon$.

Note that, up to a term of order $\epsilon$, the offset $k$ between the two lattices at the point $s_i$ is equal to the running difference $\mathcal L(s_i)$ between the two arclengths as measured from the points $\mathbf r_1(0)$ and $\mathbf r_2(0)$, respectively. Using the equations (\ref{eq:1}) and (\ref{eq:4.6}) we obtain that
\[\sum_{j=-\infty}^\infty g\left(\frac{\sqrt{{(\epsilon j+k)}^2+\delta^2}}{\delta}\right)=G(\theta(s_i),n), 
\]
where
\begin{equation}
\label{eq:6}
G(\theta,n)=\sum_{j=-\infty}^\infty g\left(\sqrt{{\left(\frac{2\pi j}{n}+\theta\right)}^2+1}\right). 
\end{equation}
Finally,
\begin{eqnarray*}
&\sum_{i=1}^{N-1}\sum_{j=1}^{N+n-1}g\left(\frac{\left|\mathbf r_{1,i}-\mathbf r_{2,j}\right|}{\delta}\right)= \sum_{i=1}^{N-1}G(\theta(s_i),n) & \\
& +o(1)=\frac{1}{\epsilon}\int_0^1G(\theta(s),n)ds+o(1)=E_w+o(1),&
\end{eqnarray*}
and the leading contribution to the total effective energy is given by a Ginzburg-Landau-type expression
\begin{equation}
\label{eq:8}
E_{eff}=E_b+E_w=\int_0^1\left(\epsilon\mu{\left(\theta^\prime\right)}^2+\frac{1}{\epsilon} G(\theta,n)\right)ds,
\end{equation}
where $\epsilon$ can be viewed as an analog of the Ginzburg-Landau parameter.

\begin{figure}[ht]
\includegraphics[width=2.5in]{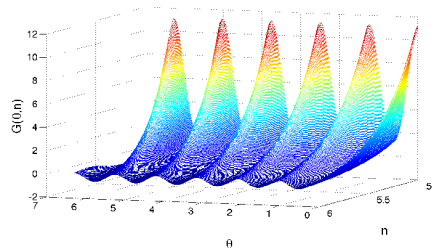}
\caption{\label{fig:Gplot} Surface of the effective potential $G(\theta,n)$.}
\end{figure}

\begin{figure}[ht]
\includegraphics[width=2.5in]{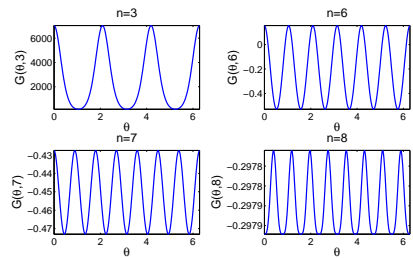}
\caption{\label{fig:Gslice} $G(\theta,n)$ for the different values of $n$.}
\end{figure}

Next we discuss the properties of the function $G$. Although $G$ is defined for all of $\theta$ and $n$, only integer values of $n$ are physically relevant as $n$ is equal to the difference between the number of  atoms on the outer curve $\mathcal C_2$ and the inner curve $\mathcal C_1$. Then, by construction, the function $G$ is ${2\pi}/{n}$-periodic in $\theta$ and it has exactly $n$ minima on the interval $[0,2\pi]$. The surface of $G$ and its crossections for various values of $n$ are shown in Figs. \ref{fig:Gplot}-\ref{fig:Gslice}.

To summarize, in our effective model a crossection of a two-walled nanotube is represented by a single curve $\mathcal C=\mathcal C_1$ with the effective energy given by (\ref{eq:8}). The equilibrium shape of the nanotube then minimizes the energy $E_{eff}$ subject to the constraints 
\begin{equation}
\label{eq:constraints}
\int_0^1\sin{\theta}\,ds=\int_0^1\cos{\theta}\,ds=0,
\end{equation}
on the function $\theta$ that enforce the closedness of $\mathcal C$. 

It is easy to see that the values of the energy minimizing function $\theta$ will reside mostly at the minima of $G(\theta,n)$ as dictated by the "penalty" term $\frac{1}{\epsilon}G(\theta,n)$.  The function $\theta$ transitions between its two constant values over the narrow "interfacial" region. The width of this region decreases with the Ginzburg-Landau parameter $\epsilon$ due to the factor of $\epsilon$ in front of the gradient term in the energy $E_{eff}$. Because $\theta$ increments by $2\pi$ as the curve $\mathcal C$ is traversed in the counterclockwise direction and because the potential function has $n$ minima on the interval $[0,2\pi]$, the minimizing configuration of $\theta$ should have exactly $n$ interfaces separating $n$ regions where $\theta$ is almost constant. Note that $n$ is equal to the number of "extra" atoms on the circumference of the outer tube and it is independent of the diameter of the inner tube (as long as the difference between the diameters of two tubes remains constant). 

Since $\epsilon=s/L$ we have that $\epsilon\to 0$ when $L\to\infty$ then the transition between the regions of constant $\theta$ must be sharper for the tubes with the larger radii. On the other hand, the interfacial regions smear out when $L$ is small and the gradient term in the energy dominates. This effect has a clear physical explanation---the energy due to the curvature-induced mismatch between the lattices of the inner and the outer tubes increases linearly with the diameter while the energy of a dislocation associated with a corner is independent of the diameter of the tube. The number of dislocations needed to remove the curvature and to introduce the appropriate lattice stacking is equal to the difference between the number of atoms along circumferences of the outer and the inner tubes. (Each dislocation "resets" the lattice to incorporate an extra atom into the outer tube.) We conclude that there must be a critical diameter above which polygonal crossections of the tubes should be energetically preferred.

\begin{figure}[ht]
\includegraphics[width=2.5in]{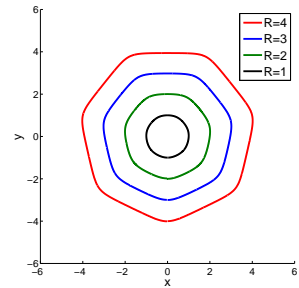}
\caption{\label{fig:polygonal} Shape of a nanotube as a function of its diameter.}
\end{figure}

If we assume that the minimizer has the symmetry \(\theta(\pi+s)=2\pi-\theta(s),\) then the constraints (\ref{eq:constraints}) are automatically satisfied and the minimizer solves the boundary value problem
\begin{equation}
\label{eq:bvp}
\left\{\begin{array}{ll}
2\mu\theta^{\prime\prime}-G_\theta(\theta,n)=0\ \mbox{if}\ 0<s<\pi, \\
\theta(0)=0\ \mbox{and}\ \theta(\pi)=\pi.
\end{array}\right.
\end{equation}
This problem was solved numerically using the MATLAB BVP solver \cite{matlab} and assuming that $\mathcal C_2$ contains seven more atoms than $\mathcal C_1$. The shapes of nanotubes of different circumferences are shown in Fig. \ref{fig:polygonal}
and correspond to the trends observed experimentally \cite{kiang}. 

Observe that in our model there is no sharp transition between the polygonal and the circular nanotube shapes and the polygonal crossections continuously morph into the circular ones as the radii of the tubes are decreased. This is the consequence of the fact that the interfaces in the "diffuse model" are not sharp---rather, the width of the interfacial regions increases continuously with $\epsilon$.

A simple geometrical calculation shows that, if two concentric circles are replaced by two concentric polygons of the same circumference, then the distance between the circles is {\em{larger}} than the distance between the polygons. However, the experimental data \cite{kiang} shows that the intershell spacing $\hat d_{002}$ {\em{decreases}} from $0.39$ to $0.34$ as the diameter of the tube increases. This is the clear consequence of turbostraticity: for large tube diameters, polygonization reduces turbostraticity for the most of the nanotube circumference and the intershell spacing decreases to the equilibrium value of the correlated stacking (the spacing can be reduced further to that of the perfect AB-stacking if a nanotube is subjected to a high-temperature heat treatment \cite{yoon}). The variation of intershell spacings for the entire range of nanotube diameters is of order $0.39-0.34=0.05$ nm---smaller than the carbon bond length of $0.142$ nm. As a consequence, the integer part of the ratio of $d/h$ remains the same for all diameters, validating the assumption that $n$ (or $d$) can be held fixed in the minimization procedure.

The extension of our approach to MWNTs in three dimensions is relatively straightforward but more technical; it will be addressed a forthcoming paper.


The authors thank A. Buldum for bringing our attention to the problem and Pat Wilber for many useful discussions. This work was supported by the NSF grant DMS 0407361.
\bibliography{nanotubes}
\end{document}